\documentclass[twocolumn,trackchanges]{aastex7} %twocolumn

\usepackage{multirow}
\usepackage{array}
\usepackage{booktabs}
\usepackage{amsmath}
\usepackage{graphicx}
\usepackage{multirow}
\newcommand{\q}[1]{\textcolor{red}{#1}}
\newcommand{\y}[1]{\textcolor{orange}{#1}}

\begin{document}

%\title{Template \aastex v7 Article with Examples\footnote{Footnotes can be added to titles}}
\title{A high-Galactic-altitude scattering screen: scintillation from Globular Cluster pulsar M53A}

\author[orcid=0009-0009-1205-7294, sname='He']{Qiuyi He}
%\altaffiliation{Kitt Peak National Observatory}
\affiliation{Department of Astronomy, Tsinghua University, 100084 Beijing, P. R. China; \href{mailto:qiuyi981022@gmail.com}{qiuyi981022@gmail.com}}
\affiliation{South-Western Institute for Astronomy Research (SWIFAR) \\ Yunnan University, 650500 Kunming, People's Republic of China; \href{mailto:xun@ynu.edu.cn}{xun@ynu.edu.cn}}
\email[show]{qiuyi981022@gmail.com}  

\author[orcid=0000-0003-2076-4510, sname='Shi']{Xun Shi} 
%\altaffiliation{Las Campanas Observatory}
\affiliation{South-Western Institute for Astronomy Research (SWIFAR) \\ Yunnan University, 650500 Kunming, People's Republic of China; \href{mailto:xun@ynu.edu.cn}{xun@ynu.edu.cn}}
\email[show]{xun@ynu.edu.cn}

\author[orcid=0000-0001-6051-3420, sname='Yin']{Dejiang Yin} 
%\altaffiliation{Las Campanas Observatory}
\affiliation{College of Physics, Guizhou University, 550025 Guiyang, People's Republic of China}
\email{..}

\author[orcid=0000-0003-3010-7661, sname='Li']{Di Li}
%\altaffiliation{Kitt Peak National Observatory}
\affiliation{Department of Astronomy, Tsinghua University, 100084 Beijing, P. R. China; \href{mailto:qiuyi981022@gmail.com}{qiuyi981022@gmail.com}}
\email{..}

%\author[orcid=0000-0003-3010-7661, sname='Li']{Di Li}
%\altaffiliation{Kitt Peak National Observatory}
%\affiliation{Department of Astronomy, Tsinghua University, 100084 Beijing, P. R. China}
%\email{..}

%\author[gname=Savannah,sname=Africa]{S. Africa}
%\affiliation{South African Astronomical Observatory}
%\affiliation{University of Cape Town, Department of Astronomy}
%\email{fakeemail3@google.com}

%\author{River Europe}
%\affiliation{University of Heidelberg}
%\email{fakeemail4@google.com}

%\author[0000-0000-0000-0003,sname=Asia,gname=Mountain]{Asia Mountain}
%\altaffiliation{Astrosat Post-Doctoral Fellow}
%\affiliation{Tata Institute of Fundamental Research, Department of Astronomy}
%\email{fakeemail5@google.com}

%\author[0000-0000-0000-0004]{Coral Australia}
%\affiliation{James Cook University, Department of Physics}
%\email{fakeemail6@google.com}

%\author[gname=IceSheet]{Penguin Antarctica}
%\affiliation{Amundsen–Scott South Pole Station}
%\email{fakeemail7@google.com}

%\collaboration{all}{The Terra Mater collaboration}

%% Use the \collaboration command to identify collaborations. This command
%% takes an optional argument that is either a number or the word "all"
%% which tells the compiler how many of the authors above the command to
%% show. For example "\collaboration[all]{(DELVE Collaboration)}" wil include
%% all the authors above this command.
%%
%% Mark off the abstract in the ``abstract'' environment. 
\begin{abstract}

%Pulsar scintillation analysis serves as an effective probe for detecting the small-scale structures and kinematic properties of the interstellar medium. 
%Currently, detected scattering screens almost exclusively concentrate near the Galactic disk, leaving the scattering properties of the deep Galactic halo as a long-standing observational blank. 
%Investigating the existence of dense discrete scattering screens in the deep halo is fundamental to reshaping our understanding of the macroscopic spatial distribution of the scattering medium and elucidating its physical nature. 
While known interstellar scattering screens are largely confined to the Galactic disk, the scattering properties 
%of the deep Galactic halo have remained unexplored.
of the circumgalactic medium (CGM) in the deep Galactic halo have remained unexplored.
Using observations from the Five-hundred-meter Aperture Spherical radio Telescope (FAST), we report the first detection of scintillation arcs in the secondary spectra of the high-latitude pulsar J1312+1810A (M53A). 
%Annual modulation modeling reveals a foreground screen at a distance of
%This study utilizes the Five-hundred-meter Aperture Spherical radio Telescope (FAST) observations of the distant globular cluster M53 at an extremely high Galactic latitude. 
%By analyzing the scintillation properties of pulsar M53A, we report the first discovery of clear scintillation arcs in its secondary spectrum. 
%Based on the annual modulation effect, we identify a foreground screen at a distance of 
%$12.8_{-3.6}^{+2.6}$\,kpc and a Galactic height of $12.6_{-3.5}^{+2.6}$\,kpc, by far the greatest altitude for any measured scintillation screen. 
Under the 1D anisotropic and 2D isotropic thin-screen models considered here, the inferred scintillation screen distances are $12.8_{-3.5}^{+2.6}$ and $11.7_{-3.5}^{+2.8}$\,kpc, respectively.
These model-dependent estimates correspond to Galactic heights of $12.6_{-3.5}^{+2.6}$ and $11.5_{-3.5}^{+2.7}$\,kpc, placing the candidate screen among the highest-altitude scattering structures inferred to date.
%\y{Correlating scintillation drift rates with dispersion measure (DM) variations indicates that screen density variations dominate the DM changes, showing column density fluctuations of $\lesssim10^{3}$\,AU\,cm$^{-3}$ on scales of $\sim10^{-3}$\,AU, suggesting an extreme column density gradient of $>10^{5}$\,cm$^{-3}$.}
The strong, time-variable scintillation asymmetry suggests substantial electron-density inhomogeneity within the screen.
%If the marginal intra-observation DM differences originate within this screen, they suggest electron-column fluctuations on transverse sampling scales of $\sim10^{-3}\,\mathrm{AU}$.}
%sub-AU scales. 
We speculate that this screen is associated with small-scale structures near the shock boundary of the northern eROSITA bubble.
%We speculate this screen is associated with the non-radiative shock shell of the northern eROSITA bubble. 
These results support the presence of discrete, inhomogeneous scattering structures deep within the Galactic halo, far beyond the regions where localized stellar feedback is typically concentrated, and highlight pulsar scintillation as a promising probe of small-scale ionized structures in the CGM.
%stellar feedback.

\end{abstract}

%% Keywords should appear after the \end{abstract} command. 
%% The AAS Journals now uses Unified Astronomy Thesaurus (UAT) concepts:
%% https://astrothesaurus.org
%% You will be asked to selected these concepts during the submission process
%% but this old "keyword" functionality is maintained in case authors want
%% to include these concepts in their preprints.
%%
%% You can use the \uat command to link your UAT concepts back its source.

%% From the front matter, we move on to the body of the paper.
%% Sections are demarcated by \section and \subsection, respectively.
%% Observe the use of the LaTeX \label
%% command after the \subsection to give a symbolic KEY to the
%% subsection for cross-referencing in a \ref command.
%% You can use LaTeX's \ref and \label commands to keep track of
%% cross-references to sections, equations, tables, and figures.
%% That way, if you change the order of any elements, LaTeX will
%% automatically renumber them.

%%\footnote{\url{http://www.latex-project.org/}}
%%\citep{lamport94} (L. Lamport 1994)
%%\citet{2015ApJ...805...23C}
%%\href{https://aas.org/posts/news/2023/08/aas-open-access-publishing-model-open-transparent-and-fair}{2023 post} 2023 post in \url.
%%\noindent :top space in a paragraph
%%\tt\string :printed letters
%%Long deluxetables should begin with the {\tt\string\startlongtable} command. This initiates a longtable environment.  Authors might have to use {\tt\string\clearpage} to isolate a long table or optimally place it within the surrounding text. http://journals.aas.org/authors/aastex.html

\section{Introduction} 
\label{sec:intro}
Multi-path interference induced by the propagation of pulsar emission through the inhomogeneous interstellar medium (ISM)---namely, interstellar scintillation \citep[ISS;][]{Rickett90}---is currently one of the few effective probes for detecting extreme small-scale structures within the ISM. 
%By analyzing parabolic scintillation arcs in secondary spectra and incorporating the annual modulation effect to break the degeneracy between distance and velocity, we can localize
%achieve precise calibration 
%of scattering screens along the loS \citep[LoS;][]{Reardon20, Huangys26}. 
Analyses of parabolic scintillation arcs and their annual modulation can constrain the locations and transverse velocities of scattering screens along the line of sight \citep[LoS;][]{Reardon20, Huangys26}. 
%Since the identified scattering structures---such as H\,II regions \citep{Mall22, Ocker24}, the Local Bubble \citep{Bhat16, Liu23}, and supernova remnants \citep{Yao21}---are predominantly concentrated near the Galactic plane \citep[typically $|{\rm Z}| < 1$\,kpc;][]{Askew23}, their physical origins are generally attributed to localized feedback activities within the stellar disk \citep{Walker17, He24}.
Many well-studied scattering structures are associated with ionized environments near the Galactic disk \citep[typically $|{\rm Z}| < 1$\,kpc;][]{Askew23}, including H\,II regions \citep{Mall22, Ocker24}, the Local Bubble \citep{Bhat16, Liu23}, and supernova remnants \citep{Yao21}. 
The physical origins of these scattering structures are therefore often attributed to localized feedback activity within the stellar disk \citep{Walker17, He24}.
%Since currently detected scattering structures are predominantly concentrated near the Galactic plane \citep[typically $|{\rm Z}| < 1$\,kpc;][]{Askew23, Ocker24}, their physical origins are widely presumed to be associated with energetic stellar feedback \citep{Walker17, He24}. 
However, does the deep Galactic halo extending into the circumgalactic medium (CGM) genuinely lack high-density fluctuating structures, or is the underlying reality merely obscured by the scarcity of bright background sources?

Predominantly located at mid-to-high Galactic latitudes and at large distances, globular cluster (GC) pulsars provide extended LoS to sample the deep halo medium \citep{He26}.
Although recent studies have successfully localized scattering screens at vertical heights of 3.0 and 4.4\,kpc toward NGC 5904 (M5) and NGC 6205 (M13) \citep{Zhangdd23}, these heights do not yet extend into the distant reaches of the deep Galactic halo.
Among GCs known to harbor pulsars, NGC 5024 (M53, $\rm {R.A. = 13^h12^m55.3^s}$, $\rm Dec = +18^{\circ}10'05''$) is the most distant, located at a heliocentric distance of 18.5\,kpc \citep{Baumgardt21}. 
Its high Galactic latitude ($b \approx 80^{\circ}$) provides an exceptionally extended LoS nearly perpendicular to the Galactic plane, sampling the Galactic halo up to a height of 18.2\,kpc.
The brightest pulsar in this cluster, J1312$+$1810A (M53A), is a 33.16\,ms millisecond pulsar residing in a wide binary system with a 256-day orbital period and extremely low eccentricity \citep{Lian25}. 
The combination of its extreme distance, high-latitude position, and relatively high flux density
%orbital stability 
renders M53A an ideal probe for detecting and characterizing scattering screens in the deep Galactic halo.

An unambiguous confirmation of strong scattering structures at extremely high Galactic latitudes could fundamentally reshape our traditional understanding of their physical origins. 
In this work, we utilize multi-epoch observations from the Five-hundred-meter Aperture Spherical Radio Telescope (FAST) to systematically analyze the interstellar scintillation characteristics of pulsar M53A in M53. 
We aim to search for discrete scattering structures in the deep Galactic halo, constrain their physical properties, and explore their possible origins.
Section \ref{sec:data} describes the observational data and data reduction procedures. 
Section \ref{sec:Results_discussion} presents and analyzes the observational results. 
Section \ref{sec:dis} discusses the physical interpretation and limitations of the inferred screen properties. 
Finally, Section \ref{sec:conclution} summarizes the findings of this work.

\section{Data and analysis} 
\label{sec:data}

We utilize the multi-epoch L-band 19-beam observational data of the globular cluster M53 obtained by FAST between 2019 and 2024 (Table~\ref{tab:M53_data}), and perform dedispersion and phase folding on the raw data using {\tt DSPSR} \citep{van11} and the latest ephemeris \citep{Lian25}.
To mitigate the persistent radio frequency interference in the FAST L-band \citep{Jiang20}, we extract the dynamic spectra from two clean subbands (1050--1150\,MHz and 1300--1450\,MHz) for independent analysis,
following similar methodology in recent studies \citep{Stinebring19, Yao21, Yao22}.
The dynamic spectra have time and frequency resolutions of 15\,s and 0.12\,MHz, respectively.
%\q{then constructing them using 15-s subintegrations while retaining the 4096 native frequency channels of the FAST data.}

\begin{table*}[t!]
    \centering
    \caption{FAST observations of M53 and derived scintillation parameters for M53A. %``Date'' and ``MJD'' denote the observing date, ``Obs.'' is the observation duration, ``Frequency'' indicates the observed band (high/low frequency), ``$\eta$'' is the scintillation arc curvature extracted from the secondary spectrum, ``$\Delta \nu_d$'' and ``$\Delta t_d$'' are the scintillation bandwidth and timescale obtained by fitting the 2D ACF slices with an exponential function, respectively, ``$\mu$'' is the drift rate, ``$\delta \mathrm{DM}/\delta t $'' is the dispersion measure derivative measured from individual observations.
    }
    \label{tab:M53_data}
    \setlength{\tabcolsep}{0pt}      
    \begin{tabular*}{\linewidth}{@{\extracolsep{\fill}}cccccccccc}
        \toprule
        Date & MJD & Obs. & Frequency & $\eta_{1.5\,\rm GHz}$ & $\Delta \nu_d$ & $\Delta t_d$ & $\mu$ & $\mu_{1.5\,\rm GHz}$ & $\delta \mathrm{DM}/\delta t $\\
       &  & (s) & (MHz) & (s$^3$) & (MHz) & (min) & (min/MHz) & (min/MHz) & (10$^{-4}$\,pc\,cm$^{-3}$\,min$^{-1}$) \\
         \midrule
\multirow{2}{*}{20191130} & \multirow{2}{*}{58817} & \multirow{2}{*}{18000} & 1300 -- 1450 & \multirow{2}{*}{$1.5_{-0.2}^{+0.3}$} & 2.34(20) & 23.01(54) & -2.77(21)& -2.13(16)& \multirow{2}{*}{0.33(42)}\\
                          &                        &                        & 1050 -- 1150 & & 2.00(12) & 18.85(57)  & -6.85(9)& -2.70(4)\\
                        \addlinespace[3pt]
\multirow{2}{*}{20210920} & \multirow{2}{*}{59477} & \multirow{2}{*}{18000} & 1300 -- 1450 & \multirow{2}{*}{--} & 1.94(16) & 12.26(38) & 0.69(30)& 0.53(23) & \multirow{2}{*}{0.19(41)}\\
                          &                        &                        & 1050 -- 1150 & & 1.19(11) & 8.15(29) & -3.88(28)&-1.53(11)\\
                          \addlinespace[3pt]
\multirow{2}{*}{20220902} & \multirow{2}{*}{59824} & \multirow{2}{*}{7200}  & 1300 -- 1450 & \multirow{2}{*}{$1.9_{-0.5}^{+0.6}$} & 2.06(15) & 19.03(66) & -5.82(14)& -4.48(11) & \multirow{2}{*}{-0.56(123)}\\
                          &                        &                        & 1050 -- 1150 & & 0.56(12) & 9.69(20) & -10.75(38)&-4.24(15)\\
                          \addlinespace[3pt]
\multirow{2}{*}{20220916} & \multirow{2}{*}{59838} & \multirow{2}{*}{7200}  & 1300 -- 1450 & -- & 1.87(33) & 14.96(406) & -2.09(26) & -1.61(20)& \multirow{2}{*}{1.11(190)}\\
                          &                        &                        & 1050 -- 1150 & $1.6_{-0.3}^{+0.3}$ & 1.38(11) & 11.51(30) & -4.78(23)&-1.89(9)\\
                          \addlinespace[3pt]
\multirow{2}{*}{20220930} & \multirow{2}{*}{59852} & \multirow{2}{*}{7200}  & 1300 -- 1450 & \multirow{2}{*}{$1.5_{-0.5}^{+0.6}$} & 3.51(11) & 18.73(86) & -1.39(7)&-1.07(5) & \multirow{2}{*}{0.13(83)}\\
                          &                        &                        & 1050 -- 1150 & & 2.04(20) & 17.38(56) & -5.96(10)&-2.35(4)\\
                          \addlinespace[3pt]
\multirow{2}{*}{20221014} & \multirow{2}{*}{59866} & \multirow{2}{*}{7200}  & 1300 -- 1450 & \multirow{2}{*}{--} & 7.10(50) & 24.06(117) & -0.32(6) &-0.25(5)& \multirow{2}{*}{1.00(166)}\\
                          &                        &                        & 1050 -- 1150 & & 2.58(7)  & 13.50(33) & -1.12(13) &-0.44(5)\\
                          \addlinespace[3pt]
\multirow{2}{*}{20230119}& \multirow{2}{*}{59963} & \multirow{2}{*}{10800} & 1300 -- 1450 & \multirow{2}{*}{--} & 9.53(13) & 27.22(44) & 0.91(10)&0.70(8) & \multirow{2}{*}{0.34(52)}\\
                          &                        &                        & 1050 -- 1150 & & 2.60(19) & 15.21(33) & -0.40(14) & -0.16(6)\\
                          \addlinespace[3pt]
\multirow{2}{*}{20230301} & \multirow{2}{*}{60004} & \multirow{2}{*}{18000} & 1300 -- 1450 & \multirow{2}{*}{--} & 5.59(5)  & 28.61(41) & 0.16(7)&0.12(5)& \multirow{2}{*}{0.17(14)}\\
                          &                        &                        & 1050 -- 1150 & & 1.86(11) & 11.80(31) & 1.23(9)&0.49(4)\\
                          \addlinespace[3pt]
\multirow{2}{*}{20230308} & \multirow{2}{*}{60011} & \multirow{2}{*}{14400} & 1300 -- 1450 & \multirow{2}{*}{$1.1_{-0.2}^{+0.4}$} & 4.79(8)  & 19.60(67) & 1.41(10) &1.09(8)& \multirow{2}{*}{-0.06(29)}\\
                          &                        &                        & 1050 -- 1150 & & 1.93(7)  & 14.30(36) & 3.61(13)&1.42(5)\\
                          \addlinespace[3pt]
\multirow{2}{*}{20230410} & \multirow{2}{*}{60044} & \multirow{2}{*}{7200}  & 1300 -- 1450 & -- & 2.38(25) & 14.39(79) & 2.16(45)&1.66(35) & \multirow{2}{*}{-0.65(180)}\\
                          &                        &                        & 1050 -- 1150 &$1.2_{-0.3}^{+0.5}$& 1.04(13) & 8.79(48) & 3.58(32)&1.41(13)\\
                          \addlinespace[3pt]
\multirow{2}{*}{20230510} & \multirow{2}{*}{60074} & \multirow{2}{*}{7200}  & 1300 -- 1450 & \multirow{2}{*}{$1.2_{-0.2}^{+0.3}$} & 3.00(20) & 18.42(63) & 2.22(21) &1.71(16)& \multirow{2}{*}{-1.79(151)}\\
                          &                        &                        & 1050 -- 1150 & & 2.08(8)  & 9.97(20) & 2.03(8)&0.80(3)\\
                          \addlinespace[3pt]
\multirow{2}{*}{20230610} & \multirow{2}{*}{60105} & \multirow{2}{*}{7200}  & 1300 -- 1450 & \multirow{2}{*}{--} & 3.93(9)  & 29.80(65) & 1.97(11)&1.52(8) & \multirow{2}{*}{-1.68(175)}\\
                          &                        &                        & 1050 -- 1150 & & 0.74(12) & 12.29(21) & 8.15(58)&3.21(23)\\
                          \addlinespace[3pt]
\multirow{2}{*}{20230710} & \multirow{2}{*}{60135} & \multirow{2}{*}{7200}  & 1300 -- 1450 & \multirow{2}{*}{--} & 4.97(15) & 33.47(74) & 2.74(14)&2.11(11)& \multirow{2}{*}{-1.60(126)}\\
                          &                        &                        & 1050 -- 1150 & & 2.89(8)  & 14.47(24) & 2.27(15)&0.90(6)\\
                          \addlinespace[3pt]
\multirow{2}{*}{20240201} & \multirow{2}{*}{60341} & \multirow{2}{*}{10800} & 1300 -- 1450 & -- & 5.76(14) & 32.39(29) & -2.31(8)&-1.78(6) & \multirow{2}{*}{0.64(68)}\\
                          &                        &                        & 1050 -- 1150 &$1.2_{-0.2}^{+0.4}$& 1.57(22) & 14.78(28) & 8.15(58) & 3.21(23)\\
                          \addlinespace[3pt]
\multirow{2}{*}{20240503} & \multirow{2}{*}{60433} & \multirow{2}{*}{10320} & 1300 -- 1450 & \multirow{2}{*}{$1.2_{-0.2}^{+0.2}$} & 2.04(10) & 16.26(70) & 4.61(13)&3.55(10)& \multirow{2}{*}{-1.47(82)}\\
                          &                        &                        & 1050 -- 1150 & & 0.78(9)  & 9.80(22) & 9.63(15)&3.80(6)\\
                          \addlinespace[3pt]
\multirow{2}{*}{20241001} & \multirow{2}{*}{60584} & \multirow{2}{*}{10800} & 1300 -- 1450 & -- & 5.02(18) & 23.50(36) & -0.74(9)&-0.57(7) & \multirow{2}{*}{0.03(31)}\\
                          &                        &                        & 1050 -- 1150 & $1.5_{-0.2}^{+0.3}$ & 3.01(12) & 20.85(21) & -3.04(9)&-1.20(4)\\
    \bottomrule
    \end{tabular*}
\end{table*}

%The SS is obtained by calculating the power of the two-dimensional Fourier transform (2D FFT) of the DS.
%We apply the NuT transform \citep{Sprenger21} to rescale the Doppler axis to 1.5\,GHz, overcoming the frequency dependence of the arc curvature. 
%The secondary spectrum is obtained using the NuT transform \citep{Sprenger21}, a generalized two-dimensional Fourier transform (2D FFT) that rescales the Doppler axis, 
We compute each secondary spectrum using a standard two-dimensional Fourier transform \citep[2D FFT;][]{Cordes06, Brisken10}
% and rescale only the resulting Doppler axis to a reference frequency of 1.5\,GHz, leaving the Fourier power unchanged.
, after which only the resulting Doppler axis is rescaled to a reference frequency of 1.5\,GHz, leaving the Fourier power unchanged.
%The secondary spectrum is \q{computed as the squared modulus of the two-dimensional Fourier transform of the dynamic spectrum using a standard fast Fourier transform (2D FFT) \citep{Cordes06, Brisken10}. 
%To compare the two subbands, we rescale the Doppler coordinate of each subband to a reference frequency of 1.5\,GHz.}
%To compare the two subbands, we rescale the Doppler coordinate of each subband to a reference frequency of $\nu_{\rm ref}=1.5\,{\rm GHz}$ using $f_{{\rm D,ref}}=f_{{\rm D,obs}}(\nu_{\rm ref}/\nu_{\rm obs})$, where $\nu_{\rm obs}$ is the central frequency of the corresponding subband.}
%The curvature $\eta$ is then measured on the rescaled Doppler coordinate.
%To compare the two subbands, we scale the Doppler coordinate to a reference frequency of 1.5\,GHz.
Due to the low signal-to-noise ratio (S/N) of M53A, we visually extract the curvature $\eta$ using reference parabolas. 
Under the thin-screen approximation \citep{Reardon20}, the curvature is related to the system parameters by:
\begin{equation}
\eta = \frac{c d_{\rm psr} s(1-s)}{2 \nu_{\rm ref}^2 V_{\rm eff}^2}.
\label{eq:eta} 
\end{equation}
%where $c$ is the speed of light, $\nu$ is the observing frequency, $d_{\rm psr}$ is the pulsar distance, $s$ is the fractional distance from the pulsar to the scattering screen, and $V_{\rm eff}$ \q{is the relevant effective velocity defined below.}
Here, $c$ is the speed of light, $\nu_{\rm ref}=1.5$\,GHz is the reference frequency, $d_{\rm psr}$ is the pulsar distance, and $s$ is the fractional distance measured from the pulsar to the screen. The relevant effective velocity $V_{\rm eff}$ depends on the adopted screen geometry.

We consider two limiting physical geometries implemented in \texttt{scintools}\footnote{\url{https://github.com/danielreardon/scintools}} \citep{Reardon20}: a 1D anisotropic screen and a 2D isotropic screen.
We define $\boldsymbol{V_{0}} = s\boldsymbol{V_{\rm obs}} + (1-s)\boldsymbol{V_{\rm psr}}$,
where $\boldsymbol{V}_{\rm obs}$ is the transverse velocity of the observer due to the Earth's orbital motion and $\boldsymbol{V}_{\rm psr}$ is the pulsar transverse velocity derived from its proper motion. 
The right-ascension and declination components of $\boldsymbol{V}_{0}$ are denoted by $V_{0,\rm RA}$ and $V_{0,\rm Dec}$, respectively.
%Their combined components along right ascension and declination are denoted by $V_{0,\rm RA}$ and $V_{0,\rm Dec}$, respectively.
For an anisotropic screen, only the velocity component parallel to the principal scattering axis contributes:
\begin{equation}
\begin{split}
V_{\rm eff}^2 =& V_{\rm eff,\parallel}^2 \\
=&
\left(
V_{0,\rm RA}\sin\zeta_1
+V_{0,\rm Dec}\cos\zeta_1
-V_{\rm scr,\parallel}
\right)^2.
\end{split}
\label{eq:anisotropic} 
\end{equation}
%where $\zeta_1$ is the orientation of the principal scattering axis projected onto the plane of the sky, $V_{\rm scr,\parallel}$ is the screen velocity projected along this axis. 
Here, $\zeta_1$ is the position angle of the principal scattering axis, and $V_{\rm scr,\parallel}$ is the screen-velocity component projected along this axis.
%In this model, $V_{\rm eff}=V_{\rm eff,\parallel}$ in Equation~\ref{eq:eta}.

For an isotropic screen, all transverse velocity components contribute, and the effective velocity is:
%\begin{equation}
%V_{\rm eff,iso}^2 =
%\left(V_{0,\rm RA}-V_{\rm scr}\sin\zeta_0\right)^2
%+\left(V_{0,\rm Dec}-V_{\rm scr}\cos\zeta_0\right)^2,
%\label{eq:isotropic}
%\end{equation}
\begin{equation}
\begin{split}
V_{\rm eff}^2 =& V_{\rm eff,iso}^2 \\
= &\left(V_{0,\rm RA} - V_{\rm scr}\sin\zeta_0\right)^2 + \left(V_{0,\rm Dec} - V_{\rm scr}\cos\zeta_0\right)^2.
\end{split}
\label{eq:isotropic} 
\end{equation}
%where $V_{\rm scr}$ is the magnitude of the screen velocity and $\zeta_0$ denotes the position angle of the screen-velocity vector.}
Here, $V_{\rm scr}$ is the magnitude of the transverse screen velocity, and $\zeta_0$ is its position angle.
%In this model, $V_{\rm eff}=V_{\rm eff,iso}$ in Equation~\ref{eq:eta}.}

We compute the two-dimensional autocorrelation function (2D ACF) for each epoch to extract scintillation scales. 
%After masking the zero-lag noise, we fit one-dimensional cuts with a single-exponential decay model to measure the scintillation bandwidth $\Delta\nu_d$​ and timescale $\Delta t_d$​ \citep{Turner24, Wang25}.
After masking zero-lag noise, we fit exponential functions to the zero-time- and zero-frequency-lag ACF cuts \citep{Turner24, Wang25}, defining $\Delta\nu_d$ as the frequency HWHM and $\Delta t_d$ as the temporal $1/e$ width, with $1\sigma$ uncertainties from the fit covariance.
To characterize the drift rate $\mu = dt/d\nu$, we fit the peak time lags across frequency-lag bins using linear least squares.

\section{Results} 
\label{sec:Results_discussion}

%\begin{figure*}[t!]
%    \centering
%	\includegraphics[width=0.49\linewidth]{figs/M53_dy_2date} 
%	\includegraphics[width=0.49\linewidth]{figs/M53_ss_2date} 
%	\caption{Dynamic spectra (left) and secondary spectra (right) of pulsar M53A for two observing epochs (20220902 and 20240503). The reversal of the tilt direction of the scintles in the dynamic spectra (which strictly corresponds to the reversal of the power asymmetry in the secondary spectra) implies a significant change in the electron density gradient along the LoS.}
%	\label{fig:M53}
%\end{figure*}

\subsection{Scintillation properties}

We extract the scintillation features of M53A from multi-epoch observations spanning five years. The dynamic spectra across all epochs exhibit extremely narrow-band structures.
%Furthermore, parabolic arcs are clearly detected for the first time in several secondary spectra \q{(Appendix~\ref{A})}. 
We detect parabolic arc power in several secondary spectra, providing the first reported scintillation-arc measurements toward M53A (Appendix~\ref{A}).
Table~\ref{tab:M53_data} summarizes the measured curvatures, alongside $\Delta \nu_d$, $\Delta t_d$, and $\mu$ extracted by fitting the 1D cuts of the 2D ACFs. 
Given that the refractive drift rate theoretically follows a steep negative power-law scaling of approximately $\mu \propto \nu^{-3}$ \citep{Gupta94, Bhat99}, we uniformly normalize the extracted $\mu$ values to a reference frequency of 1.5\,GHz for subsequent analysis.

\begin{figure*}[t!]
    \centering
	\includegraphics[width=0.98\linewidth]{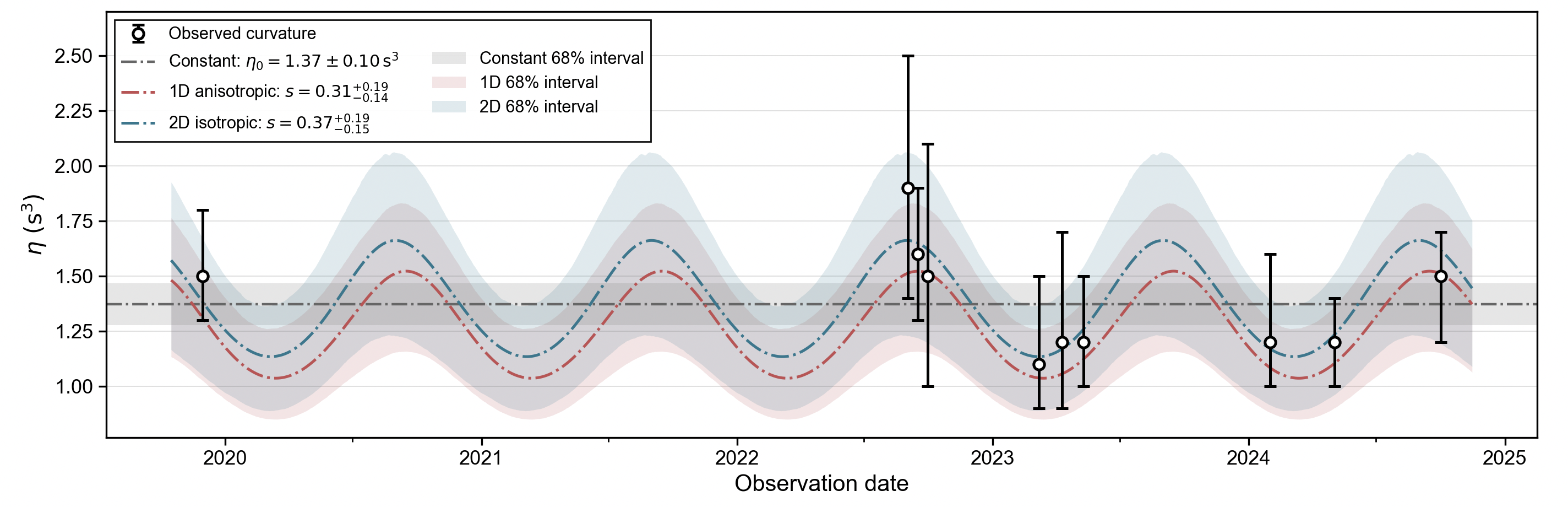}  
	\caption{Scintillation-arc curvature of M53A referenced to 1.5\,GHz. Black points show the measurements, while the gray, red, and blue curves and shaded 68\% intervals represent the constant baseline, 1D anisotropic model, and 2D isotropic model, respectively.}
	\label{fig:eta}
\end{figure*}

Observations reveal that the tilts of the scintles in the dynamic spectra exhibit significant sign reversals across different epochs. 
%This phenomenon strictly corresponds to the left-right asymmetry of the interference power along the positive and negative Doppler frequency axes ($\pm f_D$) in the secondary spectra, as illustrated by the representative observations from 2022 September 2 and 2024 May 3.
This phenomenon is qualitatively consistent with the left-right asymmetry of the interference power along the positive and negative Doppler frequency axes ($\pm f_D$) in the secondary spectra, as illustrated by the representative observations from 2022 September 2 and 2024 May 3.
%Physically, both observables may share a common origin and can be interpreted as manifestations of the large-scale refractive phase gradient within the medium, projected into different observational domains.
The pronounced asymmetry and its evolution suggest substantial electron-density inhomogeneity within the scattering screen. 
A refractive phase gradient provides a possible explanation for both the scintle drift and the asymmetric scattered image \citep{Coles10, Reardon23}.

\subsection{Scattering screen distance}

%The distance to a scattering screen can be measured from the temporal modulation of the scintillation arc curvature.
Temporal variations in scintillation arc curvature can constrain the screen distance by jointly modeling the scattering anisotropy and transverse screen velocity.
%Given the extremely low orbital eccentricity and negligible projected orbital velocity of M53A \citep{Lian25}, its scintillation arc curvature
%systemic kinematic 
%modulation is dominated by the Earth's orbital motion. 
Given the extremely low orbital eccentricity and negligible projected orbital velocity of M53A \citep{Lian25}, the Earth's orbital motion provides the dominant annual variation in the effective transverse velocity. 
As shown in Fig.~\ref{fig:eta}, $\eta$ varies only weakly relative to its measurement uncertainties, and a least-squares constant fit gives $\eta_0=1.37\pm0.10\,\mathrm{s^3}$. 
Although this curvature does not uniquely determine the screen distance, assuming pulsar-motion dominance, negligible screen velocity, and a unit projection factor gives $s\simeq0.12$, $d_{\rm scr}\simeq16.3\,\mathrm{kpc}$, and $Z\simeq16.0\,\mathrm{kpc}$ for $d_{\rm psr}\approx18.5\,\mathrm{kpc}$ \citep{Lian25}.

To relax these simplifying assumptions, we include the Earth's orbital velocity and consider both 1D anisotropic (Eq.~\ref{eq:anisotropic}) and 2D isotropic (Eq.~\ref{eq:isotropic}) screen geometries. 
We fit the corresponding annual-modulation models using \texttt{emcee} \citep{mcmc}, with parameter sets $\{s,V_{\rm scr,\parallel},\zeta_1\}$ and $\{s,V_{\rm scr},\zeta_0\}$, respectively. Their posterior distributions are presented in Appendix~\ref{B}.
The 1D anisotropic model gives $s=0.31_{-0.14}^{+0.19}$, $d_{\rm scr}=12.8_{-3.5}^{+2.6}\,\mathrm{kpc}$, and $Z=12.6_{-3.5}^{+2.6}\,\mathrm{kpc}$, whereas the 2D isotropic model gives $s=0.37_{-0.15}^{+0.19}$, $d_{\rm scr}=11.7_{-3.5}^{+2.8}\,\mathrm{kpc}$, and $Z=11.5_{-3.5}^{+2.7}\,\mathrm{kpc}$. 

%Based on the $\eta$ and the 1D anisotropic scattering screen model, we construct a dynamic curvature model that incorporates the annual modulation effect. 
%We then employ the \texttt{emcee} package \citep{mcmc} to perform global probabilistic sampling via the Markov Chain Monte Carlo (MCMC) algorithm. 
%The free parameters in our fit include $s$, $V_{\text{scr}}$, and $\zeta_1$ (the posterior distributions are presented in Appendix~\ref{B}).
%The fit yields a fractional distance parameter $s = 0.31_{-0.14}^{+0.19}$. 
%Combining this with the astrometric parameters of M53A, we precisely localize this strong scattering screen at a distance of $12.8_{-3.6}^{+2.6}$\,kpc from Earth, which corresponds to an extreme vertical height of $Z = 12.6_{-3.5}^{+2.6}$\,kpc above the Galactic plane. 
%This represents the highest and most distant scattering structure detected through pulsar scintillation to date.
The constant, 1D, and 2D fits give $\chi^2/{\rm dof}=4.94/9$, $0.685/7$, and $0.684/7$, and AICc values of 7.44, 10.68, and 10.68, respectively. 
%The sparse, low-S/N data cannot distinguish the two annual geometries. Although the constant baseline has the lowest AICc, it alone cannot constrain the screen distance; we retain both annual models for conditional distance inference. 
The constant baseline yields the lowest AICc for our sparse, low-S/N data, but it cannot constrain the screen distance. 
We therefore retain both annual geometries for conditional distance inference.
Both the physical annual models and the simplified kinematic estimate consistently indicate a large scattering distance, yielding
%Nevertheless, the conditional distance estimates are consistently large: the simplified kinematic estimate and both physical annual models give 
$d_{\rm scr}\simeq11.7$--$16.3\,\mathrm{kpc}$ and $Z\simeq11.5$--$16.0\,\mathrm{kpc}$.
Under the 1D and 2D models, the lower endpoints of the central 68\% credible intervals for $d_{\rm scr}$ are approximately 9.3 and 8.2 kpc, respectively. 
These estimates place the M53A screen substantially farther and higher than typical local screens \citep{Ocker24} and the previous screen inferred toward M13 \citep[$Z\approx4.4\,\mathrm{kpc}$;][]{Zhangdd23}, making it the highest and most distant scattering structure inferred through scintillation to date.

%Far exceeding typical local screens \citep{Ocker24} and the previous record toward M13 \citep[$Z \approx 4.2$\,kpc;][]{Zhangdd23}, our result represents the highest and most distant scattering structure detected via scintillation to date.

\subsection{Physical properties of scattering screen}

\begin{figure*}[t!]
    \centering
    \includegraphics[width=0.95\linewidth]{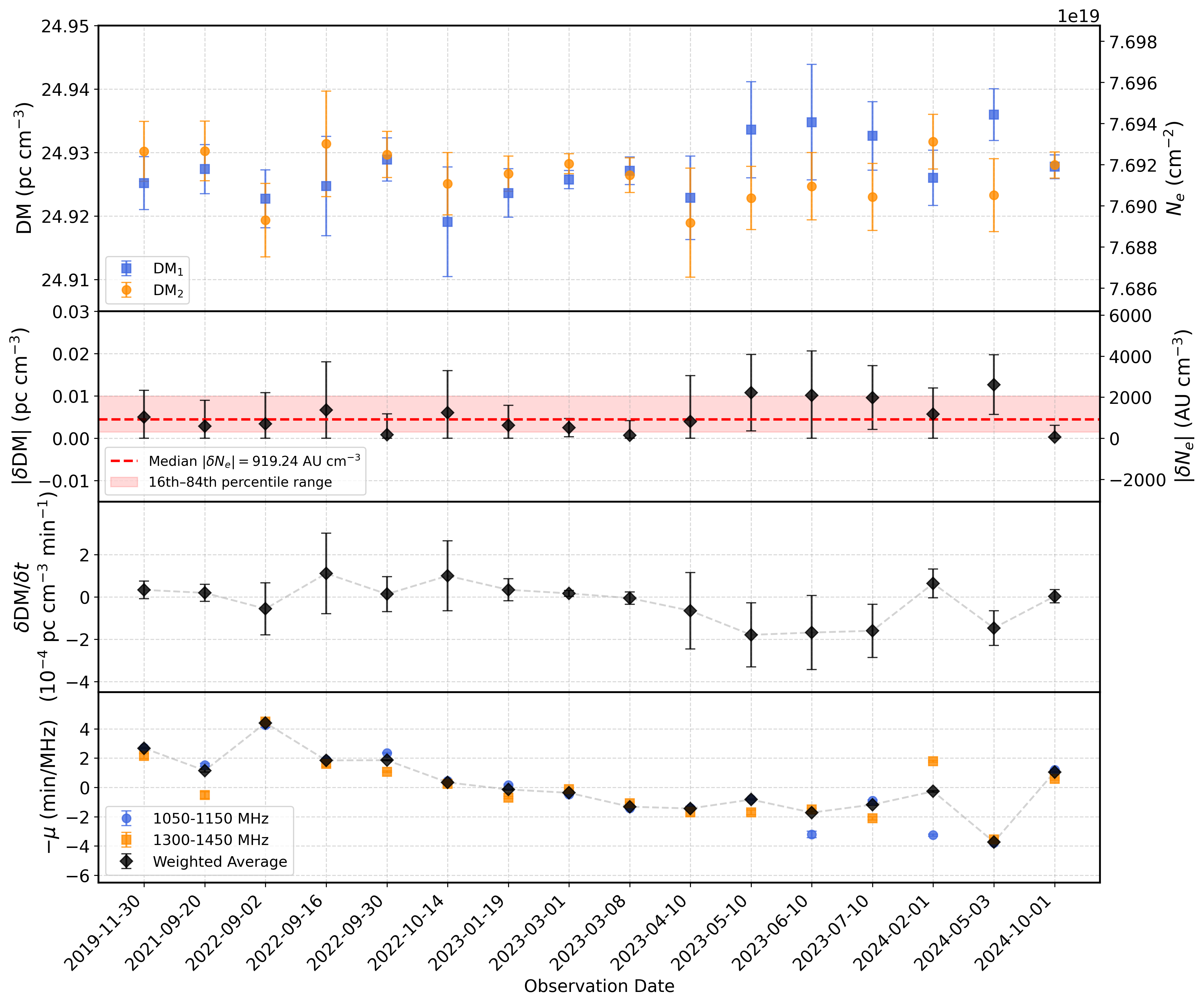}
	\caption{DM and scintillation drift variations of M53A. From top to bottom: DM measurements from the two observation halves; their absolute differences, with the median and 16th--84th percentile range shown by the red dashed line and shading; intra-observation DM variation rates; and the negative drift rate $-\mu$ referenced to 1.5 GHz for the two subbands and their weighted averages. Right-hand axes show the corresponding electron-column quantities.
	%DM variations of M53A and their correlation with the drift rate. 
	%Left: Measured values of DM$_1$ (first half) and DM$_2$ (second half) for each observation epoch (top), and their absolute difference $|\delta{\rm DM}|$ (bottom, where the red line represents the median). 
	%Right: Scatter plot of the drift rate ($-\mu$) versus the instantaneous DM variation rate ($\delta{\rm DM}/\delta t$). \q{The two quantities exhibit a positive correlation.}
	}
	\label{fig:cor}
\end{figure*}

%The extremely large physical distance to M53A ensures that the transverse scan of the LoS across the scattering screen is dominated by the pulsar's proper motion; thus, \q{the trajectory of the LoS across the scattering screen is close to a linear line.}
%no annual variation in DM is expected.
%To test whether the short-timescale dispersion measure (DM) variations originate in the scattering screen, we compare them with the drift rate, which characterizes the scintillation asymmetry. 

To further investigate electron-column variations along the LoS, we measure dispersion measure (DM) changes within individual observations. For each epoch, we estimate the within-observation DM variation rate as
\begin{equation}
\frac{\delta{\rm DM}}{\delta t} = \frac{{\rm DM}_2-{\rm DM}_1}{t_{\rm eff}/2},
\end{equation}
where ${\rm DM}_1$ and ${\rm DM}_2$ are measured from the first and second halves of the observation, respectively. Here, $t_{\rm eff}$ is the observation duration after radio-frequency interference excision.

In the thin-screen interpretation, the phase gradient projected along the effective transverse velocity, $\nabla\phi_\parallel$, shifts the scattered image and shears the 2D ACF \citep{Coles10, Reardon23}. Under the adopted geometry and sign convention, the ACF drift rate satisfies:
\begin{equation}
\mu \propto \frac{\nabla\phi_\parallel}{V_{\rm eff}}.
\end{equation}
Because $\Delta{\rm DM}\propto-\Delta\phi$, where $\Delta\phi$ is the phase change along the LoS trajectory, and $\Delta\phi=\int\nabla\phi_\parallel V_{\rm eff}\,dt$ \citep{Reardon23},
\begin{equation}
\frac{\delta{\rm DM}}{\delta t}
\propto
-\nabla\phi_\parallel V_{\rm eff}.
\end{equation}
Here, $V_{\rm eff}$ is the magnitude of the transverse velocity of the LoS relative to the screen. Combining these relations gives
\begin{equation}
-\mu\propto
\frac{1}{V_{\rm eff}^{2}}
\frac{\delta{\rm DM}}{\delta t},
\label{eq:mu_dDM_relation}
\end{equation}
assuming that the DM variations originate within the scattering screen.

The DM measurements in Figure~\ref{fig:cor} show marginal evidence for variability, with most intra-observation differences comparable to their uncertainties. These differences appear more pronounced than the changes between observing epochs. 
Furthermore, the intra-observation DM variation rates and $-\mu$ show qualitatively similar trends during some epochs, suggesting a possible connection between DM variations and scintillation asymmetry.
%, although the association remains tentative.

Under the adopted screen models, the intra-observation angular sampling separation is approximately $2\times10^{-4}\,\mathrm{mas}$, corresponding to $\sim2\times10^{-3}\,\mathrm{AU}$ at the inferred screen distance. The median observed $|\delta{\rm DM}|$ corresponds to an electron-column difference of approximately $10^{3}\,\mathrm{AU\,cm^{-3}}$. If these differences reflect physical variations dominated by the scattering screen, they imply substantial electron-column fluctuations across this sampling separation, with an estimated transverse column-density gradient of $\sim5\times10^{5}\,\mathrm{cm^{-3}}$.

\section{Discussion} 
\label{sec:dis}

%The confirmation of this scattering screen at an extreme position along the LoS to M53A poses a challenge to the traditional paradigm that attributes scattering screens predominantly to localized stellar activity within the Galactic plane.
Although the current data cannot distinguish between the 1D anisotropic and 2D isotropic geometries, both place the scattering screen at a large distance along the LoS to M53A,
 challenging the traditional paradigm that attributes scattering screens predominantly to localized stellar activity within the Galactic plane.
%to ``local stellar activity.'' 
Regardless of its specific physical origin, the extremely distant spatial distribution and intense electron density fluctuations of this structure 
provide a unique localized probe for the matter injected by large-scale Galactic dynamical processes.
%are of critical importance for understanding the dynamical evolution of large-scale Galactic structures.

\begin{figure*}[t!]
    \centering
	\includegraphics[width=0.9\linewidth]{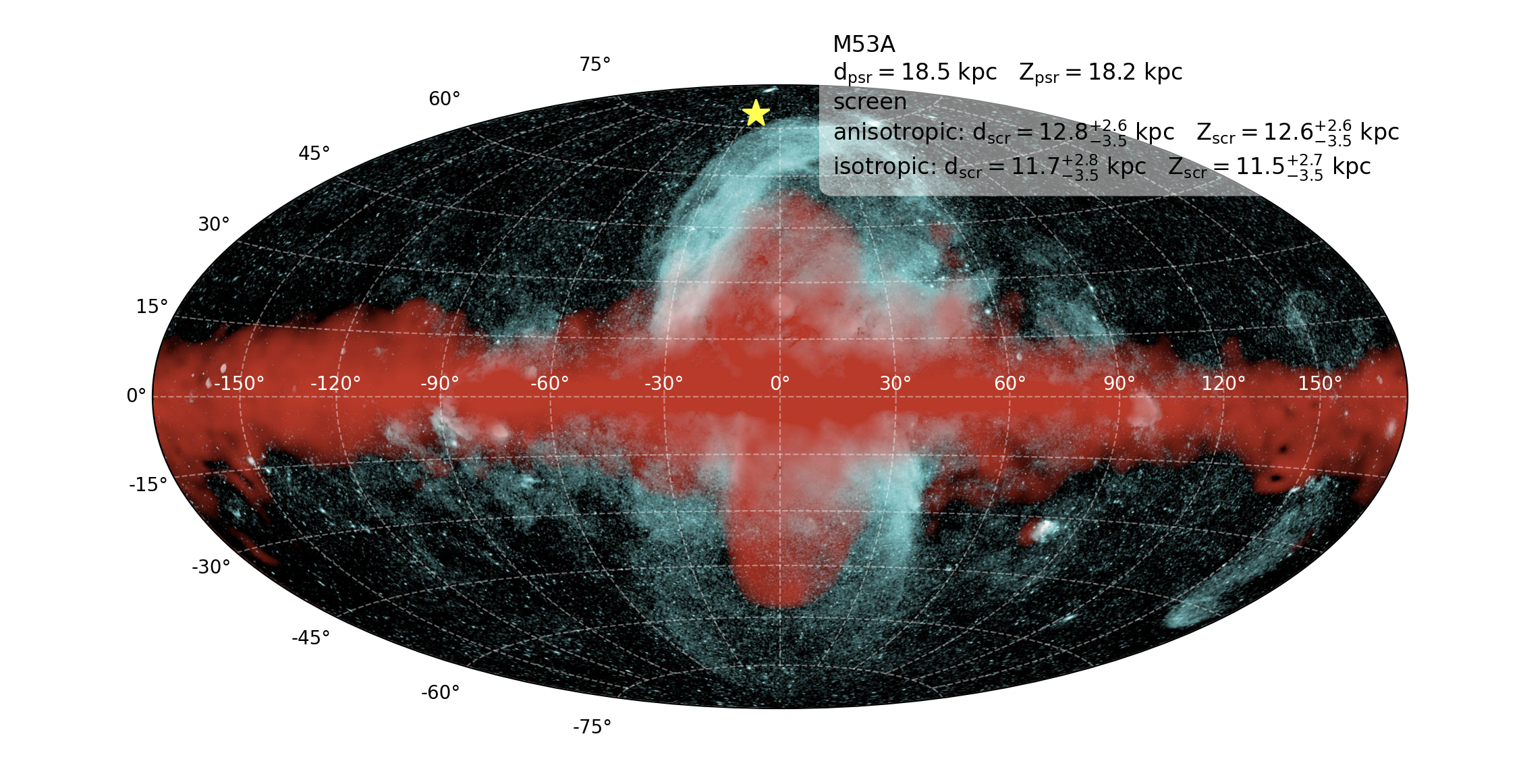} \\
	\includegraphics[width=0.8\linewidth]{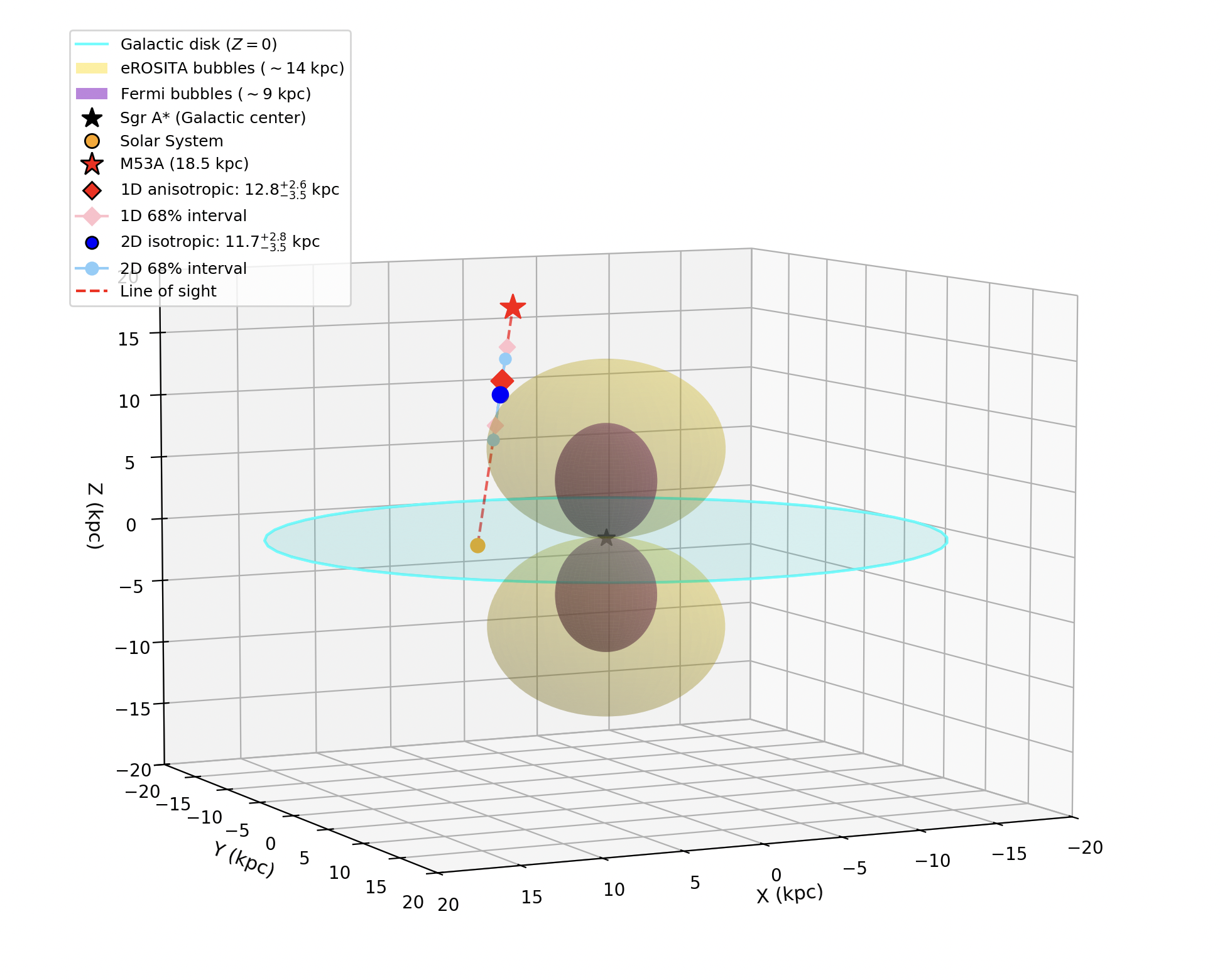}
	\caption{LoS geometry of M53A relative to Galactic bubble structures. Top: Composite all-sky map \citep[modified from][]{eROSITA} showing the Fermi bubbles (dark red, $\gamma$-ray) and eROSITA bubbles (cyan, 0.6--1.0\,keV X-ray). The yellow star marks M53A. Bottom: 3D geometric model of the M53A LoS relative to the modeled bubble structures, with dimensions based on the thick-shell model of \citet{eROSITA}.
	}
	\label{fig:eROSITA_M53A}
\end{figure*}

Based on the Spektr-RG (SRG)/eROSITA all-sky survey data, \cite{eROSITA} detected giant soft X-ray emitting bubbles (the eROSITA bubbles) extending $\sim$14\,kpc symmetrically from the Galactic center across the Galactic plane. 
Their outer boundaries manifest as non-radiative shocks formed by outward energy injection from intense starbursts or active galactic nucleus outbursts. 
Based on the 3D geometry constructed from a ``thick-shell'' model (Fig.~\ref{fig:eROSITA_M53A}), 
the height of the scattering screen determined in this study is spatially consistent with the shock-shell boundary at the top of the northern eROSITA bubble,
considering the uncertainties associated with both the X-ray boundary modeling and our fitting of the screen distance.
%The discrepancies fall well within the uncertainties associated with X-ray boundary modeling and the kinematic fitting performed.

However, if the DM differences arise within the screen, producing the inferred column-density variations over sub-AU transverse separations is difficult to reconcile with a spatially uniform, diffuse plasma. X-ray observations indicate a thermal gas density of $n_e\sim10^{-3}\,\mathrm{cm^{-3}}$ at the bubble shock boundary \citep{eROSITA}.
One possible explanation is localized overdense microstructure embedded within the macroscopic shock front.
%However, producing the observed column density variations at sub-AU scales requires a local electron density that far exceeds the extremely diffuse macroscopic thermal gas density ($n_e \sim 10^{-3}$\,cm$^{-3}$) of the bubble shock boundary derived from X-ray observations by \cite{eROSITA}.
%This discrepancy suggests that the scattering screen does not trace the bulk diffuse gas of the shock shell.
%\q{As one possible explanation,} we speculate that it corresponds to localized, overdense microstructures embedded within the macroscopic shock front.
%However, the local high-density clumps inferred from the column density variations of the scattering screen at sub-AU scales far exceed the extremely diffuse macroscopic thermal gas density ($n_e \sim 10^{-3}$\,cm$^{-3}$) of the bubble shock boundary derived from X-ray observations by \cite{eROSITA}. 
%We speculate that this scattering screen does not trace the diffuse thermal gas of the shock shell itself, \q{but rather represents thin, highly anisotropic plasma sheets or corrugated over-dense layers formed at the shock interfaces.}
%We speculate that this scattering screen does not trace the diffuse thermal gas of the shock shell itself, but rather represents the compressed remnants of pre-existing cold cloudlets in the Galactic halo swept up by the galaxy-scale shock---shock-cloud interactions can reshape such cloudlets into compact, dense substructures through hydrodynamic instabilities \citep{Klein94}. 
Such small-scale inhomogeneities may also introduce stochastic variations in $\Delta\nu_{\rm d}$ and $\Delta t_{\rm d}$, obscuring a weak annual kinematic signature.
Although difficult to resolve in X-rays, these structures can be sensitively probed through interstellar scintillation.
%Such structures are intrinsically invisible in X-ray observations, yet constitute the most sensitive targets for interstellar scintillation.

%Limited by a single LoS,
Our five-year baseline sweeps only a hundred AU across the screen, which is insufficient to geometrically constrain the macroscopic transverse extent of this structure. 
 %Nevertheless, if this structure is ultimately confirmed to be associated with the shock boundary of the eROSITA bubbles, it would demonstrate that scintillation can directly probe the complex multiphase physics of the ISM on scales inaccessible to conventional imaging.
%To verify its association with the eROSITA bubbles, 
To observationally probe the transverse extent and the properties of the discovered scattering screen,
future observations with high-sensitivity telescopes can obtain high-quality dynamic spectra for additional faint pulsars in M53 or other sources at extreme Galactic latitudes. 
By constructing a cross-sightline network of multiple targets, we can systematically test the shock-origin hypothesis,
while complementary polarimetric measurements could constrain the magnetic field along these LoSs \citep{Zhang2025SciBu}.
%and map the 3D morphology of the shock shell. 
%If confirmed as the eROSITA bubble boundary, this 
Ultimately, characterizing such extreme scattering structures
will establish pulsar scintillation as a novel probe for investigating the CGM and the evolutionary history of the Galaxy.

%Future observations with next-generation high-sensitivity telescopes hold the promise of obtaining high-quality dynamic spectra for additional faint pulsars in M53 or other sources at extreme Galactic latitudes. 
%By constructing a cross-sightline network of multiple targets, we can not only systematically test the shock-origin hypothesis but also potentially map the 3D morphology of such giant dynamical structures. 
%This will significantly expand the frontiers of pulsar scintillation techniques, establishing them as a novel probe for investigating the circumgalactic medium and the evolutionary history of the Galaxy.

\section{Conclusion}
\label{sec:conclution}

Based on multi-epoch observations spanning five years with FAST, we detected for the first time clear parabolic scintillation arcs in the secondary spectra of the extreme Galactic latitude pulsar M53A. 
%By incorporating the annual modulation of the arc curvature, we localized the scattering screen along this LoS to a Galactic height of $12.6_{-3.5}^{+2.6}$\,kpc above the Galactic plane, at a distance of $12.8_{-3.6}^{+2.6}$\,kpc. 
Conditional on the physical thin-screen interpretation, both annual modulation models yield distant-screen solutions.
%The current curvature measurements did not statistically distinguish the constant-curvature baseline from the annual descriptions. The constant baseline served only as an empirical test for resolved temporal modulation and did not provide a screen distance. 
%Conditional on the physical thin-screen interpretation, both annual modulation models yield distant-screen solutions.
The 1D anisotropic model gave $d_{\rm scr}=12.8_{-3.5}^{+2.6}$ kpc and $Z=12.6_{-3.5}^{+2.6}$ kpc, whereas the 2D isotropic model gave $d_{\rm scr}=11.7_{-3.5}^{+2.8}$ kpc and $Z=11.5_{-3.5}^{+2.7}$ kpc.
%This represents the most distant and highest-latitude discrete scattering structure precisely localized through pulsar scintillation to date, and its extreme spatial location excludes a localized origin of pulsar scintillation within the stellar disk.
This represents the most distant and highest-latitude discrete scattering structure inferred through pulsar scintillation to date, and its extreme spatial location excludes a localized origin of pulsar scintillation within the stellar disk.
%and its spatial location fundamentally excludes a local stellar origin.

%The positive correlation between $-\mu$ and the within-observation DM variation rate suggested that both shared a common physical origin, potentially driven by transverse electron-column inhomogeneities within the inferred distant scattering screen.
%On sub-AU scales,
%On scales of $\sim 10^{-3}$\,AU,
%this inhomogeneous distribution results in intense column density variations of $\lesssim10^{3}$\,AU\,cm$^{-3}$, 
%suggesting an extreme column density gradient of $>10^{5}$\,cm$^{-3}$.
%\y{The measured DM differences could indicate electron-column variations of order $10^{3}\,\mathrm{AU\,cm^{-3}}$ within the screen across transverse sampling separations of $\sim10^{-3}\,\mathrm{AU}$, corresponding to an estimated column-density gradient of $\sim5\times10^{5}\,\mathrm{cm^{-3}}$.}
The strong, time-variable scintillation asymmetry suggested substantial electron-density inhomogeneity within the screen. If the marginal intra-observation DM differences originated within the screen, they implied electron-column changes of $\sim 10^{3}\,\mathrm{AU\,cm^{-3}}$ across an inferred angular sampling separation of $\sim2\times10^{-4}\,\mathrm{mas}$.
%The evidence for an association between DM variations and scintillation asymmetry remained insufficient.
Combining the close spatial coincidence with the predicted boundary of the eROSITA bubbles and the physical picture of shock compression, we speculated that this extreme scattering screen originates from localized, overdense microstructures embedded within the non-radiative shock shell of the northern eROSITA bubble.

This research demonstrated the potential of utilizing high-Galactic-latitude pulsars to probe galaxy-scale macroscopic dynamical structures. 
In the future, as scintillation signatures of more high-latitude radio sources are confirmed, interstellar scintillation will 
%inevitably transcend its traditional role in local medium studies, 
extend beyond local medium studies, 
becoming a key probe for diagnosing the CGM and exploring Galactic evolution.

%% Please use the acknowledgment and contribution environments. This will 
%% be anonomyized when the "anonymous" style option is used. 
\begin{acknowledgments}
This work was supported by the National Natural Science Foundation of China (NSFC) under grant No. 12588202.
Additional support was provided by NSFC under grant No. 12373025.
This work was also supported by the Guizhou Provincial Natural Science Foundation project No. ZD[2026]058.
This work made use of the data from the Five-hundred-meter Aperture Spherical radio Telescope (FAST). FAST is a Chinese national mega-science facility, operated by National Astronomical Observatories, Chinese Academy of Sciences. 
We thank Zhicheng Pan for helpful discussions and assistance with the data processing of globular cluster pulsars. 
We thank the referee for the valuable comments. 
ChatGPT-5.6 and Gemini-3.1 were used to refine the manuscript to improve language clarity. All AI-assisted text was reviewed and edited by the authors, who take full responsibility for the final content.
Di Li acknowledges support from the New Cornerstone Foundation.
This version corrects the drift-rate column heading in Table 1 of the published article. This labeling correction does not affect the figures, analysis, or conclusions.
\end{acknowledgments}

\appendix
\section{Multi-epoch Dynamic and Secondary Spectra of M53A}
\label{A}

\begin{figure}[t!]
    \centering
	\includegraphics[width=\linewidth]{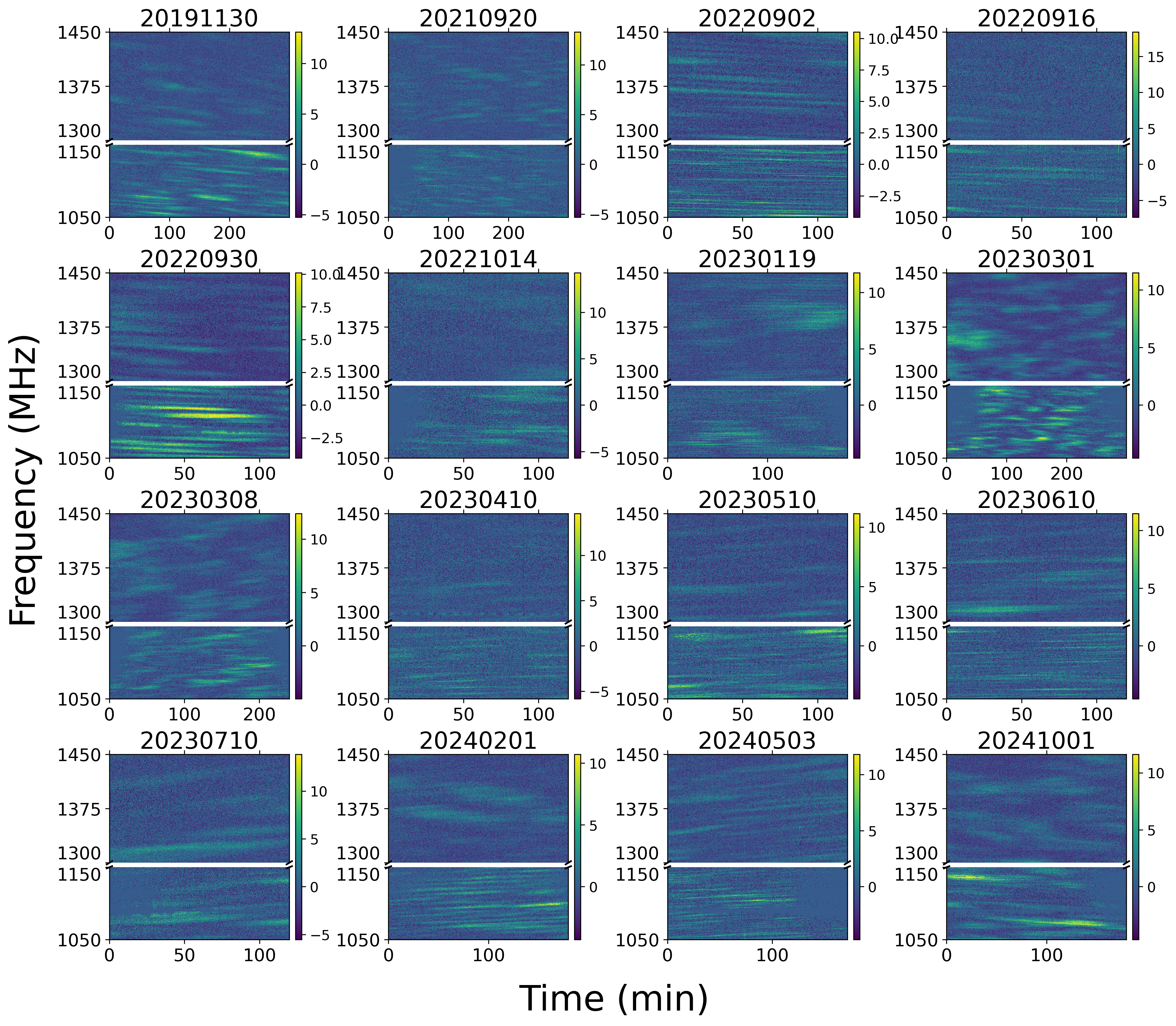} 
	\caption{Dynamic spectra of the globular cluster pulsar M53A at different observing epochs.  
	The horizontal axis represents observing time (min), and the vertical axis represents observing frequency (MHz), the color scale illustrates the normalized flux density fluctuations.}
	\label{fig:A_M53_dy}
\end{figure}

\begin{figure}[t!]
    \centering
	\includegraphics[width=\linewidth]{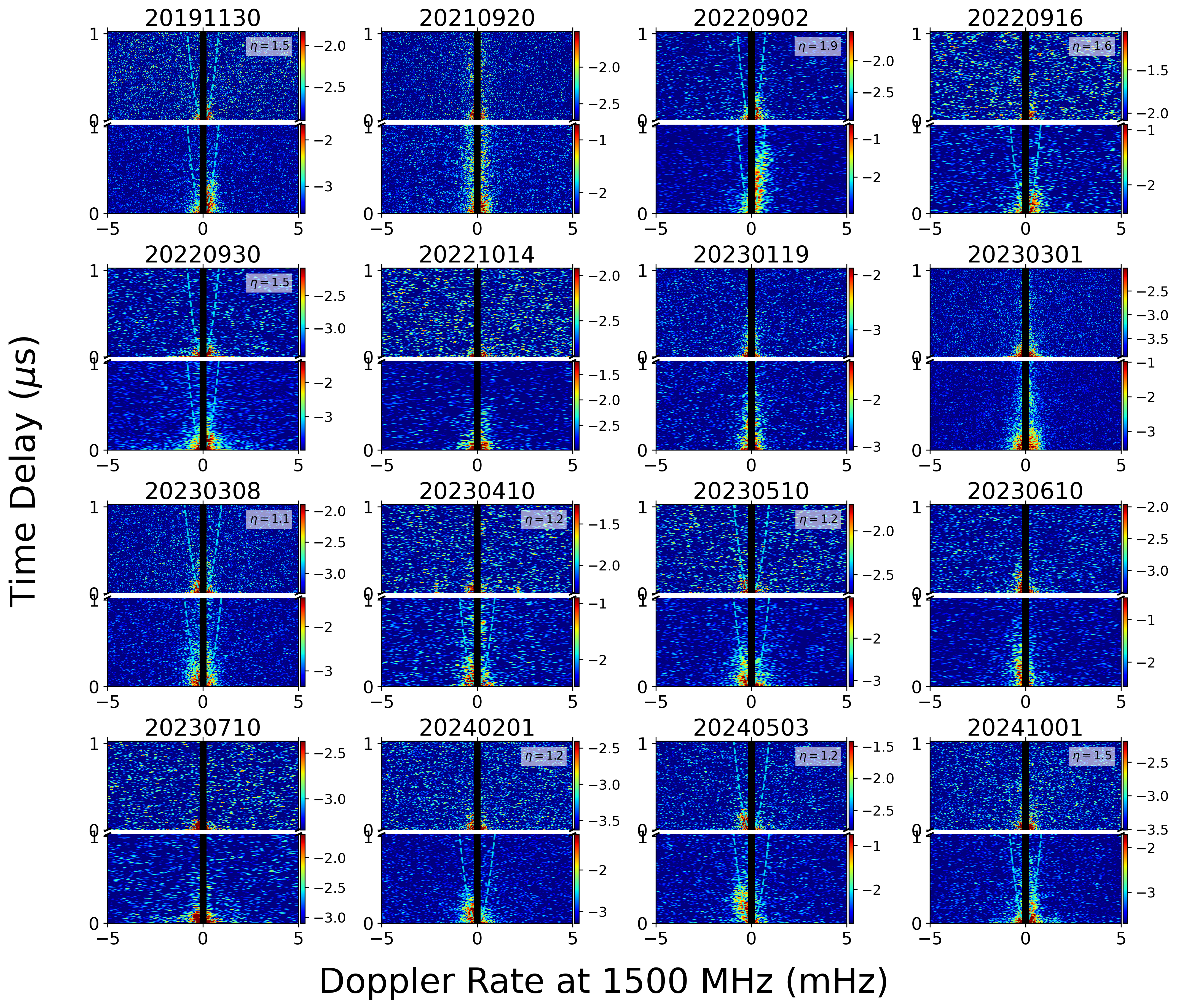} 
	\caption{Secondary spectra of the globular cluster pulsar M53A at different observing epochs.  
	Where the horizontal axis denotes the Doppler shift $f_D$ (mHz) and the vertical axis denotes the time delay $\tau$ ($\mu$s).}
	\label{fig:A_M53_ss}
\end{figure}

Figures~\ref{fig:A_M53_dy} and~\ref{fig:A_M53_ss} present the dynamic and secondary spectra of M53A for all 16 observing epochs spanning five years. 
The upper and lower panels show the high- and low-frequency subbands, respectively, with horizontal white bands marking the excised frequency ranges. 
The dynamic spectra exhibit broadly similar scintle morphologies across epochs, although their drift slopes vary. 
We overlay reference parabolas on the secondary spectra in which parabolic arcs are clearly identifiable. 
The arcs are generally diffuse, and their low S/N limits the precision of the curvature estimates, leading to broad uncertainty ranges.
The asymmetry between positive and negative Doppler power is qualitatively consistent with the direction of the scintle drift in the corresponding dynamic spectra.
No robust evidence for discrete blobs, patches, or arclets is identified, although weak features cannot be excluded because of the limited S/N.

\section{Posterior Distributions of the Screen Parameters}
\label{B}

\begin{figure}[t!]
    \centering
    \includegraphics[width=0.45\linewidth]{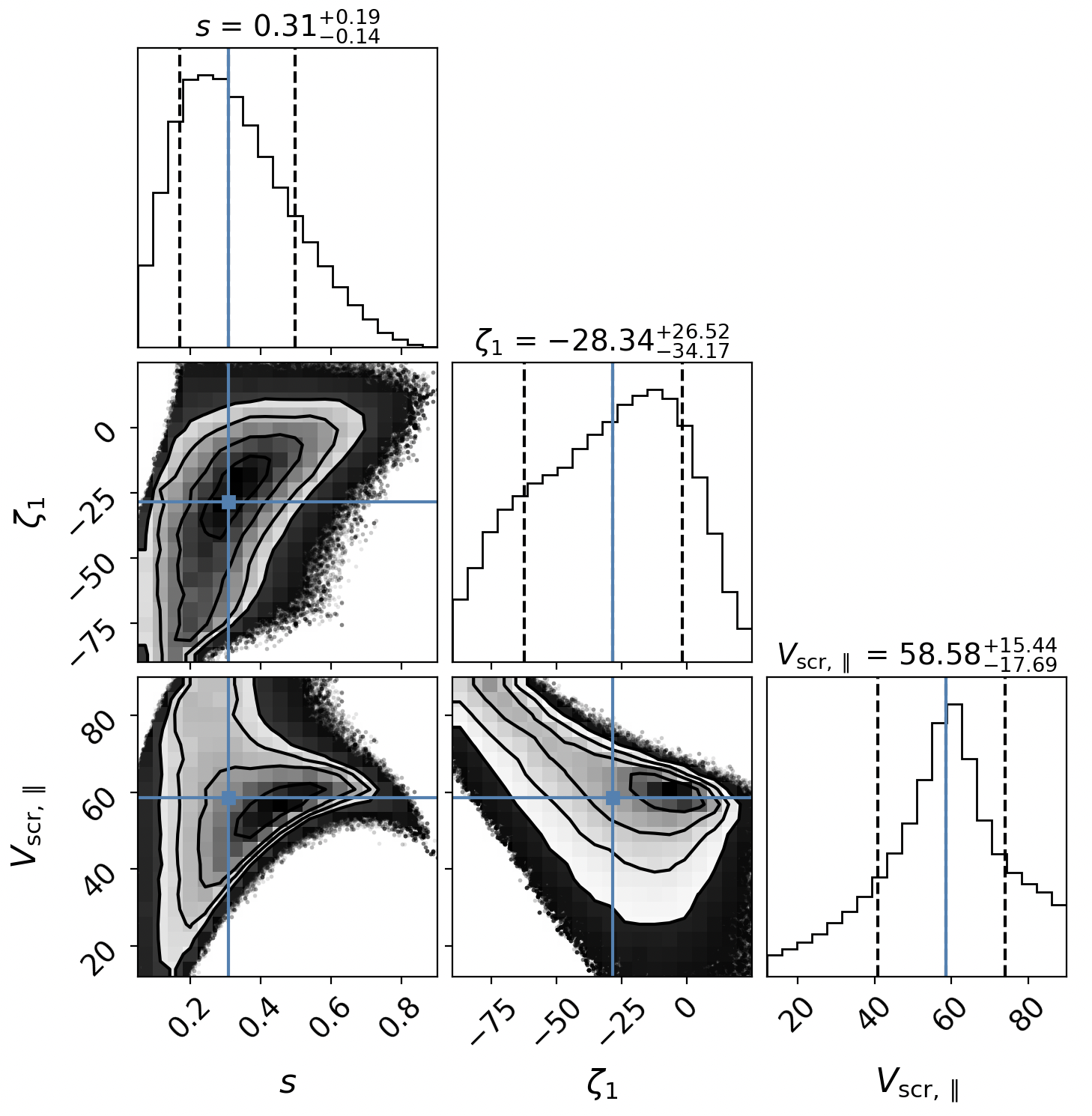}
    \hfill
    \includegraphics[width=0.45\linewidth]{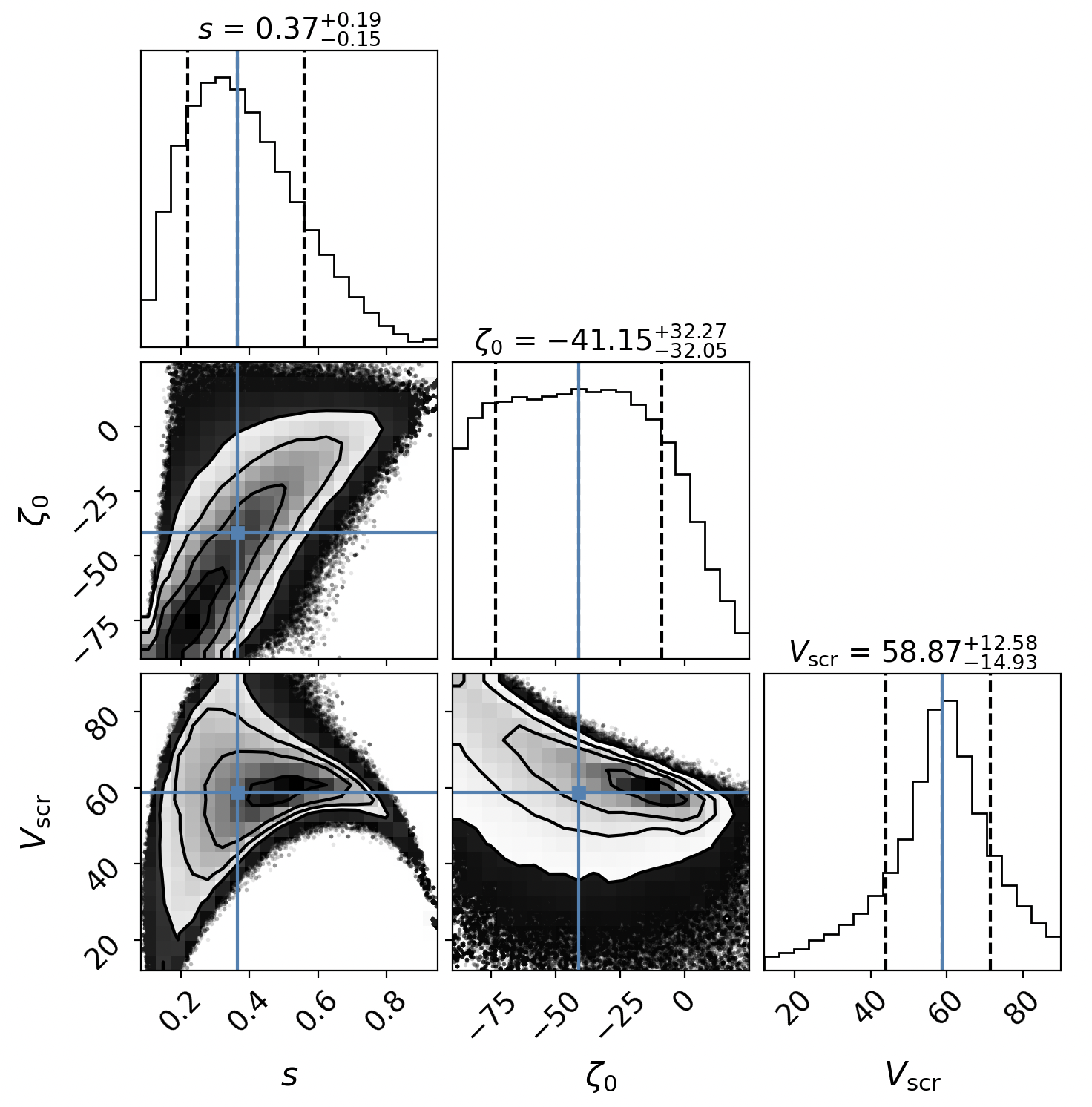}
    \caption{Posterior distributions and pairwise correlations of the annual-modulation parameters for the 1D anisotropic screen model (left) and the 2D isotropic screen model (right).}
    \label{fig:mcmc}
\end{figure}

Figure~\ref{fig:mcmc} presents the posterior distributions for the 1D anisotropic parameters $\{s,\zeta_1,V_{\rm scr,\parallel}\}$ and the 2D isotropic parameters $\{s,\zeta_0,V_{\rm scr}\}$. 
The diagonal panels show the marginalized distributions, with solid and dashed lines marking the medians and 16th--84th percentiles, respectively. 
The off-diagonal panels show the parameter correlations and nested credible regions. The broad, correlated posteriors reflect the limited constraining power of the current measurements. 
Because the arc power is concentrated near zero Doppler frequency and the S/N is low, reliable curvature measurements are available for only a subset of epochs. 
The uncertainties associated with the visual curvature measurements further broaden the posteriors. The inferred parameters remain model dependent, and the current data do not distinguish between the two screen geometries.

\bibliography{sample7}{}
\bibliographystyle{aasjournalv7}

%% This command is needed to show the entire author+affiliation list when
%% the collaboration and author truncation commands are used.  It has to
%% go at the end of the manuscript.
%\allauthors

%% Include this line if you are using the \added, \replaced, \deleted
%% commands to see a summary list of all changes at the end of the article.
%\listofchanges

\end{document}